\documentclass{article}
\usepackage{amsmath,tikz,physics,bm}
\usepackage{arxiv}
\usepackage[utf8]{inputenc} 
\usepackage[T1]{fontenc}    
\usepackage{hyperref}       
\usepackage{url}            
\usepackage{booktabs}       
\usepackage{amsfonts}       
\usepackage{nicefrac}       
\usepackage{microtype}      
\usepackage{lipsum}
\usepackage{graphicx}
\newcommand{\T}{\mathcal{T}}
  \newcommand{\E}{\mathcal{E}}
   \newcommand{\I}{\mathcal{I}}
   \newcommand{\M}{\mathcal{M}}

\title{Modeling Interfering Sources in Shared Queues for \\ Timely Computations in Edge Computing Systems}

\author{
 Nail Akar \\
  Bilkent University \\
  Bilkent, Ankara, T\"{u}rkiye\\
  \texttt{akar@ee.bilkent.edu.tr} \\
   \And
   Melih Bastopcu \\
 University of Illinois \\
Urbana-Champaign, Illinois, USA \\
\texttt{bastopcu@illinois.edu}
  \And 
 Sennur Ulukus \\
  University of Maryland\\
  College Park, MD, USA \\
  \texttt{ulukus@umd.edu} \\
  \And
  Tamer Ba\c{s}ar \\
 University of Illinois \\
Urbana-Champaign, Illinois, USA \\
\texttt{basar1@illinois.edu}
}

\begin{document}
\maketitle
\begin{abstract}
Most existing stochastic models on age of information (AoI) focus on a single shared server serving status update packets from $N>1$ sources where each packet update stream is Poisson, i.e., single-hop scenario. In the current work, we study a two-hop edge computing system for which status updates from the information sources are still Poisson but they are not immediately available at the shared edge server, but instead they need to first receive service from a transmission server dedicated to each source. For exponentially distributed and heterogeneous service times for both the dedicated servers and the edge server, and bufferless preemptive resource management, we develop an analytical model using absorbing Markov chains (AMC) for obtaining the distribution of AoI for any source in the system. Moreover, for a given tagged source, the traffic arriving at the shared server from the $N-1$ un-tagged sources, namely the interference traffic, is not Poisson any more, but is instead a Markov modulated Poisson process (MMPP) whose state space grows exponentially with $N$. Therefore, we propose to employ a model reduction technique that approximates the behavior of the MMPP interference traffic with two states only, making it possible to approximately obtain the AoI statistics even for a very large number of sources. Numerical examples are presented to validate the proposed exact and approximate models. 
\end{abstract}


\maketitle
\section{Introduction}
The advancement in IoT and 5G technologies has enabled new applications in healthcare, autonomous vehicle systems, and smart factories. For example, healthcare providers can remotely monitor the health status of their patients without requiring them to stay in the healthcare facilities. In autonomous vehicle systems, IoT devices can collect data from their surroundings to plan efficient routes. In smart factories, IoT devices can be used to predict the maintenance schedule of equipment to prevent potential delays due to machinery failures. In all these \textit{timely critical} applications, there is an increasing demand for processing data that may not be done locally on IoT devices due to their limited resources. One potential solution to enable these timely applications is to use Edge Computing (EC), which brings computation closer to real-time applications, thus reducing end-to-end communication delays \cite{mao2017survey}.

Motivated by these time-sensitive applications, the age of information (AoI) has been used to measure timeliness in communication systems \cite{yates-survey, kosta_etal_survey}. The initial work on AoI focuses on queueing networks consisting of a single source and a transmitter and uses geometric tools to characterize the average AoI at the receiver \cite{kaul_etal_ciss12, Yates__HowOftenShouldOne, najm2018status, soysal_ulukus_IT21}. The stochastic hybrid system (SHS) approach introduced in \cite{yates_mgf} and the absorbing Markov chains (AMC) approach introduced in \cite{akar_gamgam_comlet23} have enabled analysis of AoI in more involved network structures, such as gossip networks \cite{YatesGossip}, caching systems \cite{Kaswan2024_parallel_caching}, and multiple source systems \cite{yates2018age, moltafet_etal_tcom22,  cosandal2024multithresholdaoiioptimumsamplingpolicies}. Although SHS and AMC methods enable the characterization of average AoI in such networks, the computational complexity may increase exponentially, leading to impracticality in large-scale networks, thus necessitating the development of more efficient algorithms. 

In this work, we consider an EC system consisting of $N>1$ sources where each source sends its updates to an edge server in order to complete its timely computation task as shown in Fig.~\ref{fig:superposition}. When a packet is generated, sources send their packets to the edge server through their dedicated transmitter. Upon receiving packets from the source, the server (transmission or shared server) starts processing them. However, in the event that another packet arrives before the completion of the task, the current packet on the server is preempted, and the new incoming packet starts receiving service. We consider the information freshness attained in such a two-hop edge-computing system and utilize Markov modulated Poisson processes (MMPP) to simplify the state space needed to characterize the average AoI of the sources, providing a way to obtain a good approximation for the age calculation process.             

Information freshness of EC systems has been considered in \cite{Kuang2019_MEC, Ning2021_EC_health, He2024_partial, sathyavageeswaran2024timelyoffloadingmobileedge, aggarwal2024meanfieldgamemodel}. Reference \cite{Kuang2019_MEC} considers a mobile-edge-computing (MEC) problem for a single device where computations can be performed locally or on the edge server. 
The timeliness in the Internet of Medical Things (IoMT) has been studied in \cite{Ning2021_EC_health} where the computations can be done locally for IoT devices resulting in a cooperative game formulation among IoT devices or at the edge-server formulated as a non-cooperative game at the user levels with the objective of minimizing a cost related to AoI, energy consumption, and patients' health criticality. Reference \cite{He2024_partial} considers the MEC problem where each of the computation tasks can be further divided into smaller subtasks that can be completed either locally or at the edge server. The work in \cite{sathyavageeswaran2024timelyoffloadingmobileedge} considers a MEC system where the computations are done with a first-in-first-out (FIFO) queuing principle or sent to a MEC server which completes the tasks instantly. By considering Markov decision processes (MDPs), reference \cite{sathyavageeswaran2024timelyoffloadingmobileedge} finds optimal task allocations in such a system. Recently, reference \cite{aggarwal2024meanfieldgamemodel} studies the MEC problem in a non-cooperative game setting where each user's packet at the edge-server can be preempted by the others in the system. By considering a large number of devices, the authors in \cite{aggarwal2024meanfieldgamemodel} employ the mean-field game framework to find an approximate equilibrium policy for the finite-agent game problem and tackles the trade-off between the AoI and energy consumption.

In \cite{aggarwal2024meanfieldgamemodel}, the tasks can be preempted at the edge-server. However, assuming a large number of packet arrivals, in  \cite{aggarwal2024meanfieldgamemodel},  packet arrivals from the other users are approximated with a Poisson process which simplified the SHS analysis significantly at the expense of obtaining approximate age expressions for a given tagged source. Instead of approximating exogenous arrivals to the edge server with a Poisson process, in this work, we propose to utilize a two-state MMPP model that matches the first three moments of the exogenous arrival rate and also the DC spectrum of the exogenous arrivals in order to obtain a better approximation for the average AoI expressions by using the AMC method \cite{akar_gamgam_comlet23}. Through numerical results, we compare the performances of (a) a direct Poisson approximation, (b) a two-state MMPP model, and (c) the exact model that amounts to a superposition MMPP with $2^{N-1}$ states. Going beyond the EC system focused in this paper, the MMPP approximate interference modeling approach presents a novel way to reduce the state space of the AMC method which may grow exponentially with the number of users, and thereby making it possible to obtain approximate AoI statistics in networks with large number of users.

\section{Markov Modulated Poisson Process}
\label{sec:MMPP}
We provide an introduction to the Markov modulated Poisson process (MMPP) based on \cite{mmppcookbook}. An MMPP is a point process for which the rate of events taking place depends on the state of a modulating process $X(t)$, $t \geq 0$, which is an irreducible finite-state continuous-time Markov chain (CTMC). 
The modulating process $X(t)$ has an infinitesimal generator denoted by $Q$ of size $M$, called the order of the MMPP. 
When the modulating process of the MMPP visits state $m$, i.e., $X(t)=m$,  events take place according to a Poisson process with rate $r_m$. 
We define $R$ as the $M \times M$ diagonal matrix consisting of the Poisson rates in each state, 
 i.e., $R= \text{diag} \{ r_1,r_2,\ldots,r_M \}$ in which case 
 the MMPP is said to be completely characterized with the matrix pair $(Q,R)$.
Let $\pi$ be
the steady-state vector of $Q$,  
\begin{align}
\pi Q & = 0, \ \pi e=1,
\end{align} 
where $e$ is a column vector of ones of appropriate size.
The $j$th non-central moment of the rate of the MMPP characterized with the pair $(Q,R)$
is denoted by $m_j$ and is given in
\cite{mmppcookbook} as
\begin{align}
m_j = \pi R^j e, \ j \geq 1,
\end{align}
and the variance is given by $v = m_2 - m_1^2$.

We now consider the superposition of $N$ heterogeneous
independent MMPPs, which is also an MMPP whose state-space is described by Kronecker calculus \cite{bellman}. 
Given a $p \times p$ matrix $A = \{ A_{ij} \}$, and a $q \times q$ matrix $B$, the Kronecker product of the two matrices $A$ and $B$
is denoted by $A\otimes B$  which is a square matrix of size $pq$ with block elements $\{ A_{ij} B\} $.
The Kronecker sum of the matrices $A$ and $B$
is denoted by  $A\oplus B = A \otimes I_q + I_p \otimes B$, where $I_k$ denotes an identity matrix with size $k$.
The superposition of $N$ independent MMPPs, each characterized with $(Q_n,R_n), 1 \leq n \leq N$ of order $m^{(n)}$ can be represented by the superposition MMPP $(Q,R)$ of order $\prod_{n=1}^N m^{(n)}$ \cite{hellstern_tcom89},
where
\begin{align}
Q & = Q_1 \oplus Q_2 \oplus \cdots \oplus Q_{N}, \quad R  = R_1 \oplus R_2 \oplus \cdots \oplus R_{N}.
\label{superpositionprocess}
\end{align}
In case all the individual MMPPs are two-state, i.e., $m^{(n)} = 2$ for all $n$, then the superposition MMPP is of order $2^N$.

Employing MMPPs with a large state space as in \eqref{superpositionprocess} for large $N$, may lead to an analysis which is either computationally infeasible or impractical.
In this paper, we describe a technique proposed in \cite{heffes_bstj80} and used in \cite{hellstern_tcom89} that approximates an MMPP $(Q,\Lambda)$ with $m>2$ states, by a two-state MMPP which is characterized with four scalar parameters, the mean state holding times and event intensities, in each of the two states.
The proposed method of \cite{heffes_bstj80} is based on matching the first three non-central moments of the instantaneous arrival rate of the MMPP, and additionally its time constant $\tau_c$ which is defined as
\begin{align}
\tau_c & = \frac{1}{v} \int_0^{\infty} C_X(t) \dd{t}, \label{timeconstant}
\end{align}
where $C_X(t)$ is the covariance function of the arrival rate.
This time constant is also known as the DC spectrum, which was shown to be crucial for queuing performance driven by MMPP inputs \cite{Li93queueresponse}.
 In~\cite{hellstern_tcom89}, a closed-form expression is given
for $\tau_c$,
\begin{align}
\tau_c & =  \frac{1}{v} \left( \pi \Lambda (e \pi - Q)^{-1} \Lambda e - m_1^2 \right). \label{timeconstant2}
\end{align}
Let the two-state approximating MMPP be characterized with the pair $(P,\Theta)$ in the following form:
\begin{align}
P & =\left[ \begin{array}{cc}
-\sigma_1 & \sigma_1 \\
\sigma_2 & -\sigma_2 \end{array}\right], \quad \Theta=  \left[ \begin{array}{cc}
\theta_1 & 0 \\
0 & \theta_2 \end{array} \right].
\end{align}

In \cite{heffes_bstj80}, the parameters of the approximating two-state MMPP are explicitly given as
\begin{align}
\sigma_1 & = \frac{1}{\tau_c(1+\eta)}, \quad \sigma_2 = \frac{\eta}{\tau_c(1+\eta)}, \\ 
\theta_1 & = m_1 + \sqrt{v/\eta}, \quad \theta_2=m_1 - \sqrt{v \eta},
\end{align}
where
\begin{align}
\delta & =\frac{m_3-3m_1v-m_1^3}{v^{3/2}}, \quad \eta=1 + \frac{\delta}{2}(\delta - \sqrt{4+\delta^2}).
\end{align}

\section{Absorbing Markov Chain (AMC) Method for AoI Analysis}
We consider a generic status update system with $N$ information sources, $n=1,\ldots,N$, sending status update packets to a remote monitor which is tasked with timely tracking of each source.
Let us tag one of the sources, say source-$1$, for which its AoI process, i.e., time since generation of the last received packet, denoted by $\Delta_1(t)$, is illustrated in Fig.~\ref{fig:samplepath}. Let $t_j$ and $d_j$ denote the instances
at which the $j$th successful packet is generated by source-$1$ and received by the monitor, respectively. 
Unsuccessful packets are those that are generated but not received by the monitor due to preemption.
Let $s_j$ denote the system time, i.e., how long it has spent in the system after its generation, of the $j$th successful source-1 packet. Fig.~\ref{fig:samplepath} illustrates the AoI process $\Delta_{1}(t)$ (solid red thick curve) that increases with a unit slope from the value $s_j$ at time $d_{j}$ until time $d_{j+1}$.
This time interval is called cycle-$j$. The peak value in cycle-$j$ is denoted by $\Phi_{1}(j)$ which is the peak AoI (PAoI) process for source-1. Let $\Delta_n$ 
denote the steady-state random variable for the associated AoI process of source-$n$
with probability density function (pdf) $f_{\Delta_n}(x)$ and expected value $\mathbb{E} [\Delta_n]$, which 
also equals the following time average,
\begin{align}
  \mathbb{E} [\Delta_n] & = \lim_{T \rightarrow \infty} \frac{1}{T} \int_0^T \Delta_n(t) \dd{t}. 
\end{align}
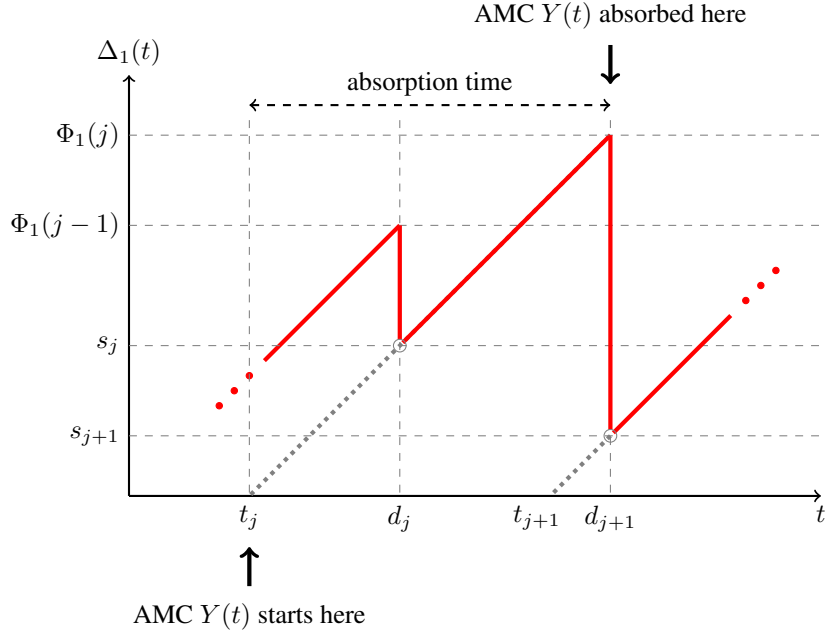
\begin{figure}[t]
	\centering
	\begin{tikzpicture}[scale=0.40]
	\draw[<->,dashed, thick,black] (4,13) -- (16,13);
	\filldraw (10,13) circle (0.0001) node[anchor=south] {absorption time};
	\draw[thick,->] (0,0) -- (23,0) node[anchor=north] {$t$};
	\draw[thick,->] (0,0) -- (0,14) node[anchor=south] {$\Delta_{1}(t)$};
	\draw[ultra thick,red] (4.5,4.5) -- (9,9);
	\filldraw[red] (4,4) circle (3pt);
	\filldraw[red] (3.5,3.5) circle (3pt) ;
	\filldraw[red] (3,3) circle (3pt); 
	\draw (0,9) node[anchor=east] {$\Phi_{1}(j-1)$};
	\draw (0,5) node[anchor=east] {$s_j$};
	\draw[dashed,gray] (0,9) -- (23,9);
	\draw[dashed,gray] (0,5) -- (23,5);
	\draw[dashed,gray] (0,2) -- (23,2);
	\draw[dashed,gray] (9,12.5) -- (9,0)  node[anchor=north, thick, black] {$d_{j}$};
	\draw[dotted,ultra thick, gray] (16,12) -- (4,0)  node[anchor=north, thick, black] {$t_{j}$};
	\draw[dotted,ultra thick, gray] (16,2) -- (14,0)  node[anchor=north, thick, black] {$t_{j+1} \quad$};
	\draw[ultra thick,red] (9,9) -- (9,5.2);
	\draw[ultra thick,red] (9.1,5.1) -- (16,12);
	\draw (0,12) node[anchor=east] {$\Phi_{1}(j)$};
	\draw (0,2) node[anchor=east] {$s_{j+1}$};
	\draw[dashed,gray] (16,12.5) -- (16,0)  node[anchor=north, thick, black] {$d_{j+1}$};
 \draw[dashed,gray] (4,12.5) -- (4,0) ;
	
	\draw[dashed,gray] (0,12) -- (23,12);
	\draw[ultra thick,red] (16,12) -- (16,2.2);
	\draw[ultra thick,red] (16.1,2.1) -- (20,6);
	\filldraw[red] (20.5,6.5) circle (3pt);
	\filldraw[red] (21,7) circle (3pt) ;
	\filldraw[red] (21.5,7.5) circle (3pt); 
	\draw[gray] (9,5) circle (6pt);
	\draw[gray] (16,2) circle (6pt);
	\draw[ultra thick,->] (4,-3) -- (4,-1.7);
	\node[black,rectangle] at (4,-4) {{AMC ${Y(t)}$ starts here}};
	\draw[ultra thick,->] (16,15) -- (16,13.7);	
	\node[black,rectangle] at (16,16) {{AMC ${Y(t)}$ absorbed here}};
	\end{tikzpicture}
	\caption{Sample path of the AoI  process for tagged source-$1$.} 
	\label{fig:samplepath}
\end{figure}

The AMC method was first introduced in \cite{akar_gamgam_comlet23} to obtain the distribution of the AoI variables $\Delta_n$ for a generic status update system, which is outlined below for tagged source-$1$.
In the first step, an AMC $Y(t)$ is constructed, which is kicked off with the generation of an arbitrary source-$1$ packet into the system, say $P_*$, which then evolves until the reception of the next successful packet from source-$1$, provided  $P_*$ is also successfully received by the monitor.
Since not all generated source-$1$ packets will eventually reach the monitor, $P_*$ may also share this fate and get thrown away from the system due to packet errors, preemption, replacement, etc. 
When this occurs, $Y(t)$ is absorbed 
into the unsuccessful absorbing state denoted by $b$. On the other hand, when $P_*$ is successfully received by the monitor, the AMC $Y(t)$ evolves until the reception of the next successful source-$1$ packet upon which absorption into the successful absorbing state, denoted by $a$, occurs. The kick-off and successful absorption instances of the AMC $Y(t)$ are shown in Fig.~\ref{fig:samplepath}.
The AMC ${Y(t)}$ then has generator ${Q}$ in the following general form, 
\begin{align}
{Q} & =\left[ \begin{array}{c|c|c} 
{S} & {s} & {u} \\ \hline
{0} & 0 & 0 
\end{array} \right], \label{AMCgenerator}
\end{align}
where the sub-generator ${S}$ represents the state transitions among the transient states, ${s}$ and ${u}$ represent the absorption rates from the transient states to states $a$ and $b$, respectively. 
In \cite{akar_gamgam_comlet23}, the following expressions are given for the pdf and the mean of AoI of source-$1$,
\begin{align}
f_{\Delta_1}(x)  & = \frac{- {\alpha} 
		\mathrm{e}^{{S}x}{h}}{{\alpha} {{S}}^{-1} {h}}, \quad
	\mathbb{E} [\Delta_1] = \frac{-{\alpha} {{S}^{-2}} {h}}{{\alpha} {S}^{-1} 
 {h}},\label{AoIEQN}
\end{align}	 
where ${\alpha}$ is the initial probability vector of the AMC $Y(t)$ that can be obtained from the steady-state solution of another recurrent MC (RMC) $Z(t)$, and ${h}$ is one for transient states visited strictly after $P_*$  is successfully received, and is zero otherwise. 
In this work, we will construct the AMC $Y(t)$ and the RMC $Z(t)$ for the specific status update system in the two-hop edge computing setting.
\begin{figure}[tb]
\centering
\begin{tikzpicture}[scale=0.30]

\draw[  thick,->] (0,0) -- (4,0) ;
 \draw[  thick](6,0) circle (2);
 \draw (6,0) node[] {{$\mu_N$}} ;
\draw (2,0) node[anchor=south] {$\lambda_N$};

\draw (2,3) node[anchor=south] {$\vdots$};
\draw (6,3) node[anchor=south] {$\vdots$};
\draw (10,5) node[anchor=south] {$\vdots$};
\draw[  thick,->] (0,8) -- (4,8) ;
 \draw[  thick](6,8) circle (2);
 \draw (6,8) node[] {{$\mu_3$}} ;
\draw (2,8) node[anchor=south] {$\lambda_3$};

\draw[  thick,->] (0,13) -- (4,13) ;
 \draw[  thick](6,13) circle (2);
 \draw (6,13) node[] {{$\mu_2$}} ;
\draw (2,13) node[anchor=south] {$\lambda_2$};
\draw[  thick,->] (8,13) -- (13,13) ;
\filldraw[color=red] (10.5,13) circle (0.3) node[anchor=center] {};
\draw[  thick,->] (8,8) -- (13,11.5) ;
\filldraw[color=red] (10.5,9.75) circle (0.3) node[anchor=center] {};
\draw[  thick,->] (8,0) -- (14.7,10.5) ;
\filldraw[color=red] (11.35,5.25) circle (0.3) node[anchor=center] {};

\draw[  thick,->] (0,21) -- (4,21) ;
 \draw[  thick](6,21) circle (2);
 \draw (6,21) node[] {{$\mu_1$}} ;
\draw (2,21) node[anchor=south] {$\lambda_1$};
\draw[ thick,->] (8,21) -- (13,21) ;
\draw[  thick](15,21) circle (2);
\draw (15,21) node[] {{$+$}} ;
\draw[ thick,->] (17,21) -- (26,21) ;
\filldraw (32,22) node[anchor=center] {\small{shared server}};
\filldraw (32,20) node[anchor=center] {\small{$\sigma$}};
\draw[  thick](15,13) circle (2);
\draw (15,13) node[] {{$+$}} ;
\draw[ thick,->] (15,15) -- (15,19) ;
\filldraw[color=blue] (15,17) circle (0.4) node[anchor=center] {};
\filldraw (15.5,18)  node[anchor=west] {\small{interference from} };
\filldraw (16.5,16)  node[anchor=west] {\small{sources $2$ to $N$}};

\draw[rounded corners,thick] (26,19) rectangle (38,23) {};

\end{tikzpicture}
\caption{From the perspective of source-1, status update system with $N$ sources with Poisson packet arrivals $\lambda_n$ from source-$n$, $N$ dedicated preemptive transmission servers with service rate $\mu_n$ for source-$n$, and one shared server with service rate $\sigma$.}
\label{fig:superposition}
\end{figure}
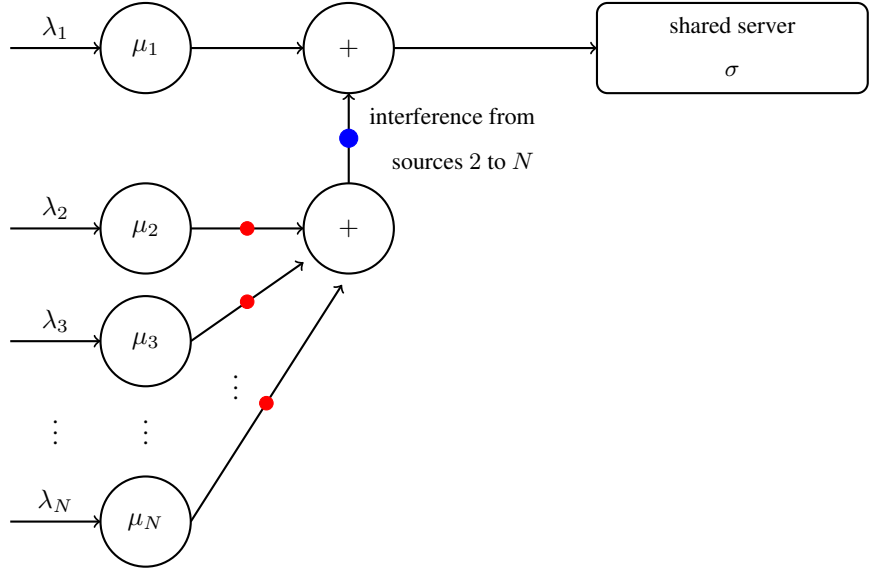

\section{System Model}
The status update system of interest is illustrated in Fig.~\ref{fig:superposition}.
Each source $n=1,\ldots,N$ generates status update packets according to a Poisson process with intensity $\lambda_n$. The status update packets from source-$n$ first receive service from 
a dedicated bufferless preemptive transmission server with exponentially distributed  service times with rate $\mu_i$. Dedicated servers represent the transmission links from each source to the edge computing server. 
Departures from each dedicated server, i.e., completed transmissions, then receive service from the edge computing server which is also bufferless and fully preemptive with exponentially distributed service time with parameter $\sigma$. Service completion epochs at the edge computing server represent packet receptions by the monitor.

Let us look at the system from the point of view of source-$1$ in Fig.~\ref{fig:superposition}.
Departures from transmission server $n=2,\ldots,N$ (shown by the red circles) is a two-state MMPP $(P_n,\Theta_n)$, where
\begin{align}
P_n & =\left[ \begin{array}{cc}
-\lambda_n & \lambda_n \\
\mu_n & -\mu_n \end{array}\right],  \quad \Theta_n= \left[ \begin{array}{cc}
0 & 0 \\
0 & \mu_n \end{array} \right],
\end{align}
since there are no packet departures when the server is idle (state 1) and there are packet departures with rate $\mu_i$ when the server is busy (state 2). 
From the perspective of source-$1$, we propose three alternatives to model the superposition traffic at the superposition point (blue circle) which are:
\begin{itemize}
    \item Method 1: Poisson approximation that matches only the first moment of the arrival rate.
    \item Method 2: Two-state MMPP model using the model reduction method of \cite{heffes_bstj80} that matches the first three moments of the arrival rate and also the DC spectrum, which was detailed in Section~\ref{sec:MMPP}.
    \item Method 3:
     Exact superposition MMPP model with $2^{N-1}$ states as given in \eqref{superpositionprocess}.
\end{itemize}
In order to cover all the three methods above, in the rest of the paper, we assume that the interference traffic from the sources 2 to $N$ is modeled with a general MMPP characterized with the matrix pair $(Q,R)$ of order $M$, $Q =  \{ q_{i,j} \}, i,j \in \M = \{1,2,\ldots,M\}$ and $R=\text{diag} \{ r_1, r_2,\ldots,r_M \}$.
Hence, Methods 1, 2, and 3 correspond to the choices of $M$ being $1$, $2$, and $2^{N-1}$, respectively.

\section{Analytical Method}	
The AMC-based method we propose for the two-hop  edge computing system is based on \cite{akar_gamgam_comlet23} and comprises the following two steps. We observe from Fig.~\ref{fig:samplepath} that, a single AoI cycle (e.g., cycle-$j$) starts with the reception of a successful packet (e.g., at time $d_j$) and continues until the reception of the next successful packet (e.g., at $d_{j+1})$. In the first step, we construct an AMC $Y(t)$ with two absorbing states, which starts operation at time $t=0$ (corresponding to $t=t_j$ in Fig.~\ref{fig:samplepath}) with the generation of $P_{\ast}$. The transient and absorbing states (resp.~transition rates) of this AMC are given in Table~\ref{tab:AMC} (resp.~Table~\ref{tab:TRofAMC}). 
Note that there are $8M$ transient states for the AMC $Y(t)$.
In order to understand the operation of $Y(t)$, notice that when $P_{\ast}$ is successful, then the AMC continues evolving until the reception of the next successful packet upon which we reach the successful absorbing state $a$ (corresponding to $t=d_{j+1}$ in Fig.~\ref{fig:samplepath}). If this packet is discarded or preempted, the AMC is absorbed into the unsuccessful absorbing state $b$.
In these tables, ${\T}$ stands for dedicated transmission server for source-$1$ and $\E$ stands for 
the edge computing server, whereas
$\I$ refers to the interference MMPP process.
The binary state of server $\T$ is state 0 (server is idle) or state 1 (ongoing transmission). Similarly, server $\E$ is in state 0 (server is idle or carrying source-$n$ packets, $n \neq 1$) or in state 1 (ongoing transmission from source-$1$). At this point, through the use of the transition rates in Table~\ref{tab:TRofAMC}, we obtain the sub-generator ${S}$ of size $8M$ as in \eqref{AMCgenerator} which governs the transitions among the transient states.  The column vector ${h}$ is one corresponding to the states $(m,i,j), m \in\M, i,j=0,1$ which are visited strictly after $P_*$ is successfully received. The only parameter that is missing in \eqref{AoIEQN}  is the initial probability vector ${\alpha}$ for which purpose we construct an RMC $Z(t)$
which has only the states $(m,i,j), m \in\M, i,j=0,1$ with the purpose of finding the status of the status update system at the arrival epoch of $P_*$. The transition rates of this RMC are given in Table~\ref{tab:TRofRMC}.
Let $\pi(m,i,j)$ be the steady-state probability of being in state $(m,i,j)$ at an arbitrary time which also equals the same probability at the arrival epoch of $P_*$ due to the PASTA property \cite{gross_etal_book}.
We observe that the packet $P_*$ will start its journey in the AMC $Y(t)$ in its state $(m,j)_{\T}$ with probability $\pi(m,0,j)+\pi(m,1,j)$, which concludes the construction of the initial probability vector ${\alpha}$.
With ${\alpha},{S},$ and ${h}$ constructed with the procedure described above, the equations in \eqref{AoIEQN} provide the distribution of AoI for source-$1$ and in particular its mean $\mathbb{E}[\Delta_1]$.
Clearly, this procedure can be repeated for each source-$n$, $2 \leq n \leq N$ to obtain the distributions of AoI for the other sources as well.


\begin{table}[tb]
\caption{Transient and absorbing states of the AMC process $Y(t)$ constructed for source-$1$.}
\centering
\begin{tabular}{|c|c|} 
 \hline
 State & Description \\
 \hline \hline
 $(m,j)_{\T}, m \in \M, j=0,1$ & $P_{\ast}$ on $\T$, $\I$ in $m$, $\E$ in $j$ \\
 \hline
 $(m,i)_{\E}, m \in \M, i=0,1$ & $P_{\ast}$ on $\E$, $\I$ in $m$, $\T$ in $i$  \\
 \hline
 $(m,i,j), m \in\M, i,j=0,1$ & $\I$ in $m$, $\T$ in $i$, $\E$ in $j$ \\
 \hline
 $a$ & successful absorbing state \\ \hline
 $b$ & unsuccessful absorbing state \\ \hline 
\end{tabular}
\label{tab:AMC}
\end{table}

\begin{table}[htb]
\caption{Transition rates for the AMC process $Y(t)$.}
\centering
\begin{tabular}{|c|c|c|} 
 \hline
 \multicolumn{3}{|c|}{Transition Rates}  \\
 \hline \hline
 From & To  & Rate \\ 
 \hline \hline
 $(m,j)_{\T}$ &  $(m',j)_{\T}$ when $m' \neq m$ & $q_{m,m'}$  \\ \cline{2-3} 
   & {\color{black}$(m,0)_{\T}$ when $j=1$} & $r_m$  \\ \cline{2-3} 
   & $(m,0)_{\T}$  & $\sigma$  \\ \cline{2-3}  
 & $(m,0)_{\E}$ & $\mu_1$  \\ \cline{2-3} 
 & $b$ & $\lambda_1$ \\ \hline
$(m,i)_{\E}$ &  $(m,i,0)$  & $\sigma$  \\ \cline{2-3} 
   & $(m,1)_{\E}$ when $i=0$ & $\lambda_1$  \\ \cline{2-3} 
   & $(m',i)_{\E}$ when $m' \neq m$ & $q_{m,m'}$  \\ \cline{2-3}  
 & $b$ when $i=1$ & $\mu_1$  \\ \cline{2-3} 
 & $b$ & $r_m$ \\ \hline
 $(m,i,j)$ &  $(m',i,j)$ when $m' \neq m$ & $q_{m,m'}$  \\ \cline{2-3} 
   & $(m,i,0)$ when $j=1$ & $r_m$  \\ \cline{2-3} 
   & $(m,1,j)$ when $i=0$ & $\lambda_1$  \\ \cline{2-3}  
 & $(m,0,1)$ when $i=1$ & $\mu_1$  \\ \cline{2-3} 
 & $a$ & $\sigma$ \\ \hline
   
\end{tabular}
\label{tab:TRofAMC}
\end{table}

\begin{table}[htb]
\caption{Transition rates for the RMC process $Z(t)$.}
\centering
\begin{tabular}{|c|c|c|} 
 \hline
 \multicolumn{3}{|c|}{Transition Rates}  \\
 \hline \hline
 From & To  & Rate \\ 
 \hline \hline

 $(m,i,j)$ &  $(m',i,j)$ when $m' \neq m$ & $q_{m,m'}$  \\ \cline{2-3} 
   & $(m,i,0)$ when $j=1$ & $r_m + \sigma$  \\ \cline{2-3} 
   & $(m,1,j)$ when $i=0$ & $\lambda_1$  \\ \cline{2-3}  
 & $(m,0,1)$ when $i=1$ & $\mu_1$  \\ \hline
\end{tabular}
\label{tab:TRofRMC}
\end{table}

\section{Numerical Examples}
\begin{figure}[t]
		\centering
	\includegraphics[width=0.6\columnwidth]{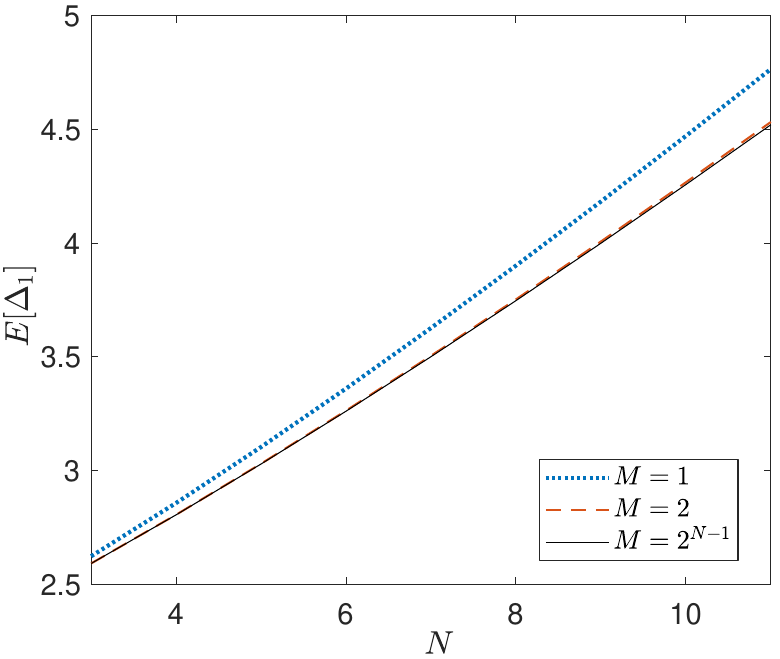}
 
 \caption{$\mathbb{E}[\Delta_1]$ depicted as a function of $N$ for interference MMPP modeling methods 1-3 when $\delta=0.2$.}
 \label{example1}\vspace{-0.35cm}
	\end{figure}

\begin{figure}[t]
		\centering
	\includegraphics[width=0.60\columnwidth]{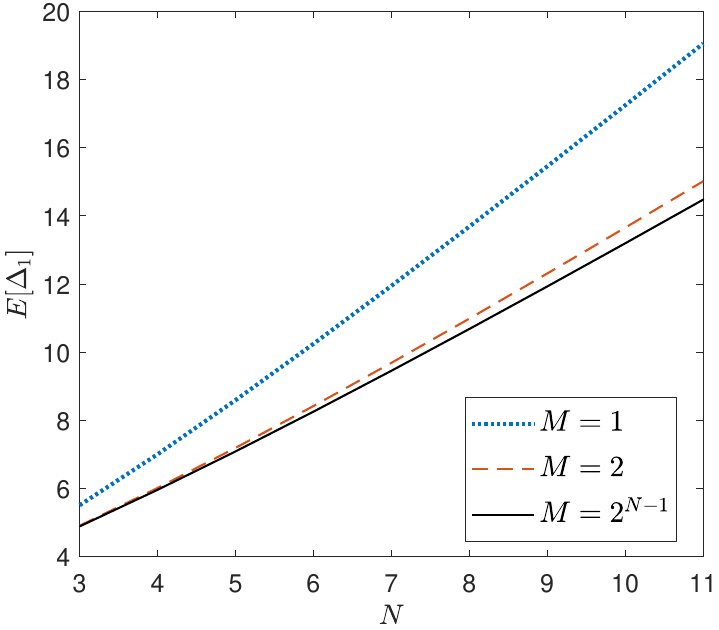}
  \caption{$\mathbb{E}[\Delta_1]$ depicted as a function of $N$ for interference MMPP modeling methods 1-3 when $\delta=1$.}
   \label{example2}\vspace{-0.4cm}
	\end{figure}

In the numerical examples, we vary the number of sources $N$ from $3$ to 11 and we fix 
$\lambda_n =1$, $\forall n$, and the service rate of the edge server as $\sigma=5$.  The service rates of the transmission servers are linearly spaced as $\mu_1=1, \ \mu_n = \mu_{n-1} + \delta, n=2,\ldots,N$ where the parameter $\delta$ represents the level of heterogeneity among the transmission servers, with heterogeneity increasing with increased $\delta$.
The mean AoI for source-$1$ is depicted as a function of the number of sources $N$ in Fig.~\ref{example1} for $\delta=0.2$ and in Fig.~\ref{example2} for $\delta=1$. The Poisson approximation for interference traffic works poorly for both examples. The two-state MMPP model reduction technique is very effective in modeling the interference traffic with the approximation being more accurate for $\delta=0.2$, which is a relatively less heterogeneous scenario than the case $\delta=1$.

\section{Conclusion}
In this work, we have considered timeliness in an edge computing system consisting of $N>1$ heterogeneous sources that send their computing tasks to the edge server through their local transmitter. In this system, an arriving task can preempt the current packets both at the local transmitter and at the edge server. Due to the preemption at the local transmitter, the arrival process from the other (i.e., interfering) sources at the edge server is no longer Poisson. In such an edge computing system with multiple sources, we noted that the complexity of finding the exact average AoI increases exponentially with the number of sources. For that, we utilized two-state Markov Modulated Point Process (MMPP) to find an approximate average AoI expression efficiently. Going beyond the edge computing problem focused on this work, the presented MMPP approach can be used to simplify the state-space required to find AoI in such networks with multiple sources.




\appendix









\end{document}